\documentclass{article}
\usepackage{mlspconf}
\usepackage{cite}
\usepackage{amsmath,amssymb,amsfonts}
\usepackage{algorithmic}
\usepackage{graphicx}
\usepackage{textcomp}
\usepackage{xcolor}
\usepackage{subcaption}
\title{SELF-ATTENTION FOR AUDIO SUPER-RESOLUTION}
\name{Nathanaël Carraz Rakotonirina}
\address{Université d'Antananarivo, Madagascar}

\begin{document}

\maketitle

\begin{abstract}
Convolutions operate only locally, thus failing to model global interactions. Self-attention is, however, able to learn representations that capture long-range dependencies in sequences. We propose a network architecture for audio super-resolution that combines convolution and self-attention. Attention-based Feature-Wise Linear Modulation (AFiLM) uses self-attention mechanism instead of recurrent neural networks to modulate the activations of the convolutional model. Extensive experiments show that our model outperforms existing approaches on standard benchmarks. Moreover, it allows for more parallelization resulting in significantly faster training.

\end{abstract}
\section{Introduction}
\label{sec:intro}

Audio super-resolution is the task of reconstructing a high-resolution audio signal from a low-resolution one. The sampling rate of the low-resolution signal is increased. It is also known as time series super-resolution \cite{birnbaum2019temporal} or bandwidth extension \cite{ekstrand2002bandwidth}. Since audio signals are long sequences, long-range dependencies need to be captured in order to obtain good performance. The architectures that are used to process sequential inputs include convolutional neural networks (CNN) \cite{lecun1989backpropagation} and recurrent neural networks (RNN) \cite{elman1990finding, hochreiter1997long}. Recent audio super-resolution models use deep neural network (DNN) with dense layers \cite{li2015dnn} or one-dimensional convolutions \cite{kuleshov2017audio}. Temporal Feature-Wise Linear Modulation (TFiLM) \cite{birnbaum2019temporal} uses recurrent neural networks to modify the activations of the convolutions. However, there are limitations associated with convolutions and RNNs due respectively to limited receptive fields and the vanishing gradient problem \cite{bengio1994learning}. In contrast, self-attention \cite{vaswani2017attention}, a recent advance for sequence modeling and generative modeling tasks, is able to capture long-range information and is more parallelizable. 

In this paper, we consider the use of the self-attention mechanism for the audio super-resolution task. The architecture we propose makes use of a combinations of self-attention and convolution. We introduce the Attention-based Feature-Wise Linear Modulation (AFiLM) layer which relies on self-attention instead of recurrent neural networks to alter the activations of the convolutions. It allows the model to efficiently incorporate long-range information. We evaluate the model on the VCTK \cite{yamagishi2012english} and Piano \cite{mehri2016samplernn} datasets. Experiments show that our approach outperforms previous methods.

\begin{figure}
    \centering
    \begin{subfigure}[b]{0.15\textwidth}
        \centering
        \includegraphics[width=\textwidth]{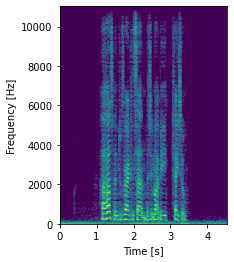}
        \caption{High-resolution}
    \end{subfigure}
    \begin{subfigure}[b]{0.15\textwidth}
        \centering
        \includegraphics[width=\textwidth]{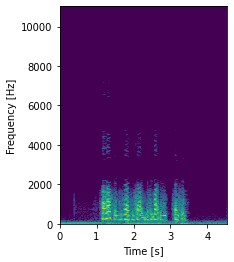}
        \caption{Low-resolution}
    \end{subfigure}
    \begin{subfigure}[b]{0.15\textwidth}
        \centering
        \includegraphics[width=\textwidth]{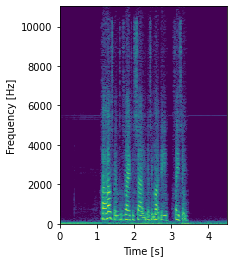}
        \caption{Reconstruction}
    \end{subfigure}
    \caption{Spectrograms of the high-resolution signal, the subsampled low-resolution signal and the reconstructed signal.}
    \label{fig:spectrogram}
\end{figure}

\section{Related work}
\label{sec:related_work}
\paragraph*{Audio super-resolution.}
Audio super-resolution, also known as bandwidth extension \cite{ekstrand2002bandwidth,larsen2005audio}, is the task of predicting high-resolution signal using a low-resolution one. It predicts the signal’s high frequencies from its low frequencies. More formally, given a low-resolution audio signal $x=(x_{1/R_1},...,x_{R_1T/R_1})$ which is sampled at a rate $R_1/T$, we want to reconstruct a high-resolution version $y=(y_{1/R_2},...,y_{R_2T/R_2})$ of $x$ with a sampling rate $R_2 > R_1$. We note $r=R_2/R_1$ the upscaling factor. Spectrograms are presented in Figure \ref{fig:spectrogram} to illustrate it.

Early audio super-resolution models utilize matrix factorization \cite{bansal2005bandwidth,liang2013beta}. They are trained on very small datasets due to the computational cost of factorizing matrices. Dong et al. \cite{dong2015audio} use analysis dictionary learning. Learning-based methods exploit Gaussian mixture models \cite{cheng1994statistical,pulakka2011speech,park2000narrowband} and linear predictive coding \cite{bradbury2000linear}. Li et al. \cite{li2015dnn} introduce a neural network with dense layers. The first convolutional architecture is proposed by Kuleshov et al. \cite{kuleshov2017audio} to scale better with dataset size. Time-Frequency Network (TFNet) \cite{8462049} works in both the time and frequency domain. Wang et al.\cite{wang2018speech} builds on the WaveNet model \cite{oord2016wavenet}. Macartney et al. \cite{stoller2018wave} use the  Wave-U-Net  \cite{macartney2018improved} architecture for audio super-resolution. Some approaches \cite{eskimez2019speech,hu2020phase,li2018speech,li2019speech,kumar2020nu} leverage  generative adversarial networks (GANs) \cite{goodfellow2014generative}. Temporal Feature-Wise Linear Modulation (TFiLM) \cite{birnbaum2019temporal} uses RNNs to alter the activations of the convolutional layers. Our approach is based on TFiLM but we use self-attention instead of RNNs.

\paragraph*{Self-attention.}
Self-attention produces a weighted average of values computed from hidden units using a similarity function. It has been used for sequence modeling because of its ability to capture long-range interactions \cite{bahdanau2014neural, bello2016neural}. The Transformer architecture \cite{vaswani2017attention} using only attention without any other architecture produces state-of-the-art results in Machine Translation. Recent speech enhancement approaches \cite{koizumi2020speech,hao2019attention,giri2019attention} also use self-attention. Attention augmented convolutions \cite{bello2019attention} achieve significant improvement in discriminative visual tasks. This combination of convolution and
self-attention is also applied to speech recognition \cite{gulati2020conformer}. Our approach is similar in that it captures both local and global dependencies of an audio sequence.

\begin{figure}[htb]
    \centering
    \includegraphics[width=.4\textwidth]{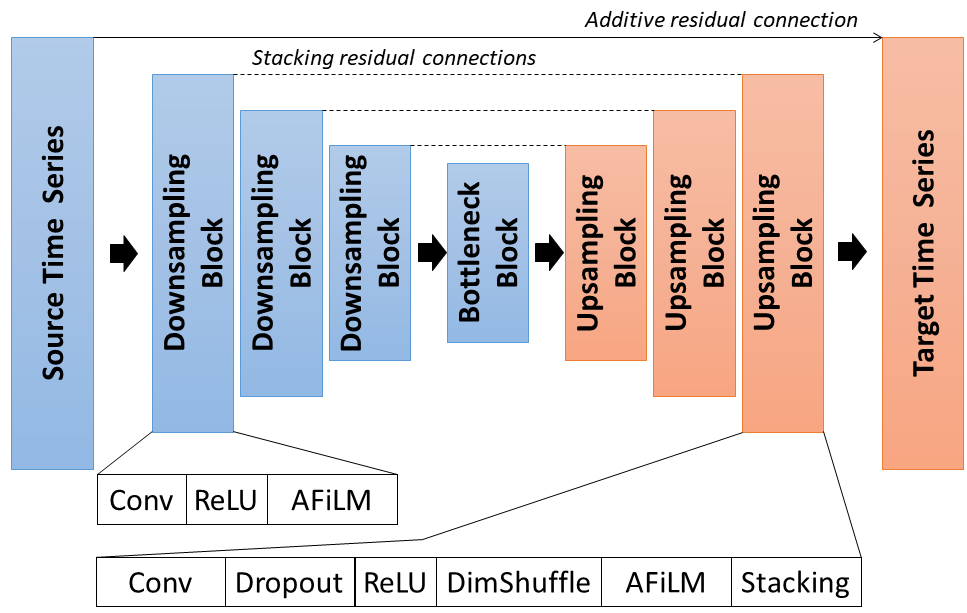}
    \caption{Network architecture used for audio super-resolution where $K=4$ as used in our implementation. It consists of downsampling blocks, a bottleneck block and upsampling blocks.}
    \label{fig:arch}
\end{figure}

\section{Proposed method} 
\label{sec:afilm}

\subsection{Network architecture} 
\label{subsec:arch}
We use the following naming conventions : $T$ and $C$ are, respectively, the 1D spatial dimension and the number of channels. The network architecture, presented in Figure \ref{fig:arch}, follows the overall architecture of \cite{kuleshov2017audio, birnbaum2019temporal} which is composed of $K$ downsampling blocks, a bootleneck layer and $K$ upsampling blocks; there are symmetric residual skip connections \cite{he2016deep} between blocks. The bottleneck is the same as the downsampling blocks but with Dropout \cite{srivastava2014dropout}. Each downsampling block $k=1,2,...,K$ contains $\max(2^{6+k},512)$ filters of length $\min(2^{7-k}+1,9)$ with a stride of $2$. Each upsampling block $k=1,2,...,K$ contains $\max(2^{7+(K-k+1)},512)$ of length $\min(2^{7-(K-k+1)}+1,9)$ .In the upsampling blocks, a one-dimensional version of the subpixel shuffling layer \cite{shi2016real} is used for upscaling.

\subsection{Attention-based Feature-Wise Linear Modulation layer} 
\label{subsec:afilm}
We present a novel neural network layer that exploits the self-attention mechanism in order to capture long-range dependencies. Feature-Wise Linear Modulation (FiLM) \cite{perez2017film} is a neural network component that applies an affine transformation to feature maps conditioned on some input. A function called FiLM generator outputs the normalizers $(\gamma,\beta)$ which modulate the activations of a neural network called  FiLM-ed network. In Temporal Feature-Wise Linear Modulation (TFiLM) \cite{birnbaum2019temporal}, the FiLM-ed network is a convolutional model and the FiLM generator network is an RNN. This is a self-conditioned model since the inputs of the FiLM generator are the activations themselves. However, RNNs are less effective in modeling very long sequences \cite{bengio1994learning}. Besides, their sequential nature can be computationally limiting. On the other hand, self-attention can represent dependencies regardless of their distance in the input sequence.

\begin{figure*}[htb]
    \centering
    \includegraphics[width=\textwidth]{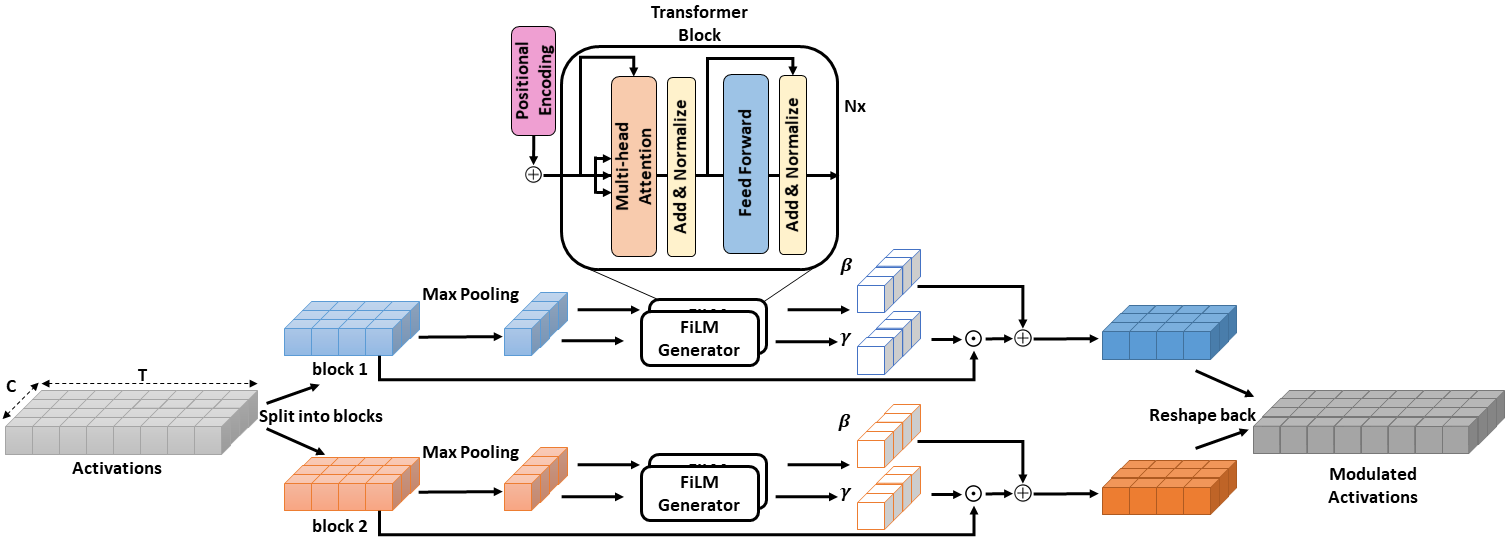}
    \caption{The AFiLM layer uses a Transformer block as a FiLM generator. Element-wise multiplication and addition are performed to modulate the feature maps. This illustration uses $T=8, C=4, B=2$.}
    \label{fig:atfilm}
\end{figure*}

We propose the Attention-based Feature-Wise Linear Modulation (AFiLM) layer which makes use of self-attention as the FiLM generator to modulate the activations of the convolutional model as depicted in Figure \ref{fig:atfilm}. Our FiLM genetator is the Transformer's \cite{vaswani2017attention} basic block. Following TFiLM, the feature map modulation is applied on a block level. The tensor of activations $F \in \mathbb{R}^{T \times C}$ is split into $B$ blocks resulting in a block tensor $F^{block} \in \mathbb{R}^{B \times T/B \times C}$. Max pooling is then used along the second dimension to downsample $F^{block}$ into a tensor $F^{pool} \in \mathbb{R}^{B \times C}$. The normalizers $\gamma, \beta \in \mathbb{R}^{B \times C}$ are then obtained by applying the FiLM generator. For each block $b$, the parameters are computed as follows:
\begin{equation}
    (\gamma[b,:], \beta[b,:]) = \textrm{TransformerBlock}(F^{pool}[b,:])
\end{equation}
where TransformerBlock is composed of a stack of multi-head self-attention and point-wise fully
connected layers. Residual connections and layer normalization \cite{ba2016layer} are used in the sub-layers. After that, for each block $b$, the normalizers $\gamma_b$ and $\beta_b$ modulate the activations via a feature-wise affine transformation:
\begin{equation}
    \textrm{AFiLM}(F^{block}[b,t,c]) = \gamma[b,c] \cdot F^{block}[b,t,c]  + \beta[b,c]
\end{equation}
Finally, the resulting tensor is reshaped back into its original shape $(T,C)$. Different combinations of $\gamma$ and $\beta$ can modulate feature maps in multiple ways. Independently of the length of the feature map, AFiLM can efficiently influence it taking into account the global interactions. Furthermore, by replacing the RNNs with self-attention, the model benefits from better parallelization which significantly speeds up training.  The AFiLM layer is added at the end of each downsampling, bottleneck and upsampling block as presented in Figure \ref{fig:arch}. Our implementation uses $K=4$ and a Transformer block that is composed of a stack of 4 layers with the number of heads $h=8$ and hidden dimension $d=2048$. 

\section{Experiments}
\label{sec:experiments}
\subsection{Datasets}
The model is trained and evaluated on the VCTK \cite{yamagishi2012english} and Piano \cite{mehri2016samplernn} datasets. The VCTK dataset contains speech data from 109 native speakers of English with various accents. Each speaker reads out about 400 diferent sentences. There is a total of 44 hours of speech data. The Piano dataset contains 10 hours of Beethoven sonatas. Both datasets are used at a sampling rate of 16 kHz. Following previous works \cite{kuleshov2017audio,birnbaum2019temporal}, we apply an order 8 Chebyshev type I low-pass filter before subsampling the high-resolution signal. The single speaker task trains the model on
the first 223 recordings of VCTK Speaker 1 which is approximately 30 minutes and tests on the last 8 recordings. Concerning the multi speaker task, we train on the first 99 VCTK speakers and test on the 8 remaining ones. The Piano dataset is split into
88\% training, 6\% validation, and 6\% testing.

\subsection{Training details}
Our model is trained  for 50 epochs on patches of length 8192, as are existing audio super-resolution models. This ensures a fair comparison. The low-resolution audio signals are first processed with bicubic upscaling before they are fed into the model. The learning-rate is set to $3 \times 10^{-4}$. The model is optimized using Adam \cite{kingma2014adam} with $\beta_1=0.9$ and
$\beta_2 = 0.999$. 

\begin{table}[h]
\centering
\caption{Quantitative evaluation of audio super-resolution models at different upsampling rates. Left/right results are SNR/LSD (higher is better for SNR while lower is better for LSD). Baseline results are those reported in \cite{birnbaum2019temporal}.}
\label{tab:evaluation}
\vspace{1em}
    \begin{tabular}{|l|r|r|r|r|}
    \hline
    Method & \multicolumn{1}{c|}{Scale} & \multicolumn{1}{c|}{\begin{tabular}[c]{@{}c@{}}VCTK\\ Single\end{tabular}} & \multicolumn{1}{c|}{\begin{tabular}[c]{@{}c@{}}VCTK\\ Multi\end{tabular}} & \multicolumn{1}{c|}{Piano} \\ \hline
    Bicubic & 2                          & 19.0/3.5                                                                     & 18.0/2.9                                                                    & 24.8/1.8                   \\
    DNN \cite{li2015dnn}   & 2                          & 19.0/3.0                                                                     & 17.9/2.5                                                                    & 24.7/2.5                   \\
    CNN \cite{kuleshov2017audio}    & 2                          & 19.4/2.6                                                                     & 18.1/1.9                                                                    & 25.3/2.0                   \\
    TFiLM \cite{birnbaum2019temporal} & 2                          & \textbf{19.5}/2.5                                                                     & 19.8/1.8                                                                    & 25.4/2.0                   \\
    AFILM   & 2                          & 19.3\textbf{/2.3}                                                                          & \textbf{20.0/1.7}                                                                         & \textbf{25.7/1.5}                        \\ \hline
    Bicubic & 4                          & 15.6/5.6                                                                     & 13.2/5.2                                                                    & 18.6/2.8                   \\
    DNN \cite{li2015dnn}   & 4                          & 15.6/4.0                                                                     & 13.3/3.9                                                                    & 18.6/3.2                   \\
    CNN \cite{kuleshov2017audio}      & 4                          & 16.4/3.7                                                                     & 13.1/3.1                                                                    & 18.8/2.3                   \\
    TFiLM \cite{birnbaum2019temporal} & 4                          & 16.8/3.5                                                                     & 15.0/2.7                                                                    & 19.3/2.2                   \\
    AFILM   & 4                          & \textbf{17.2/3.1}                                                                          & \textbf{15.4/2.3}                                                                         & \textbf{20.4/2.1}                        \\ \hline
    Bicubic & 8                          & 12.2/7.2                                                                     & 9.8/6.8                                                                     & 10.7/4.0                   \\
    DNN \cite{li2015dnn}   & 8                          & 12.3/4.7                                                                     & 9.8/4.6                                                                     & 10.7/3.5                   \\
    CNN \cite{kuleshov2017audio}     & 8                          & 12.7/4.2                                                                     & 9.9/4.3                                                                     & 11.1/2.7                   \\
    TFiLM \cite{birnbaum2019temporal} & 8                          & 12.9/4.3                                                                     & 12.0/2.9                                                                    & \textbf{13.3}/2.6                   \\
    AFILM   & 8                          & \textbf{12.9/3.7}                                                                          & \textbf{12.0/2.7}                                                                         & 12.9\textbf{/2.5}                        \\ \hline
    \end{tabular}
\end{table}

\subsection{Results}
The metrics used to evaluate the model are the the signal to noise ratio (SNR) and the log-spectral distance (LSD) \cite{gray1976distance}. These are standard metrics used in the signal processing literature. Given a reference signal y and the corresponding approximation x, the SNR is
defined as
\begin{equation}
    \textrm{SNR}(x,y) = 10\log\frac{ ||y||_2^2 }{||x-y||_2^2}
\end{equation}
The LSD measures the reconstruction quality of individual
frequencies and is defined as
\begin{equation}
    \textrm{LSD}(x,y)= \frac{1}{T}\sum_{t=1}^{T} \sqrt{\frac{1}{K}\sum_{k=1}^{K} \bigg( X(t,k) - \hat{X}(t,k) \bigg)^2}
\end{equation}
where $X$ and $\hat{X}$ are the log-spectral power magnitudes of $y$ and $x$ defined as $X = \log |S|^2$, where S is the short-time Fourier transform (STFT) of the signal. $t$ and $k$ are respectively index frames and frequencies.

We compare our best models with existing approaches at upscaling ratios 2, 4 and 8. The results are presented in Table \ref{tab:evaluation}. All the baseline results are those reported in \cite{birnbaum2019temporal}. Our contributions result in an average improvement of 0.2 dB over the TFiLM in terms of SNR. Concerning the LSD metric, our approach improves by 0.3 dB on average. This shows that our model effectively uses the self-attention mechanism to capture long-term information in the audio signal. Our attention-based model outperforms all previous models on the multi speaker task. It is the most difficult task and is the one that benefits most from more long-term information.

\begin{table}[]
\centering
\caption{Training time evaluation. The number of seconds per epoch was obtained using an NVIDIA Tesla K80 GPU.}
\label{tab:computational_per}
\begin{tabular}{|l|c|c|}
\hline
Model                & TFiLM  & AFiLM  \\ \hline
Number of parameters & 6.82e7 & 1.34e8 \\ \hline
Seconds per epoch    & 370    & 276    \\ \hline
\end{tabular}
\end{table}

\begin{table}[h]
\centering
\caption{Out-of-distribution evaluation of the model. The model is trained on the VCTK Multispeaker and Piano datasets at scale $r=2$ and tested both on the same and on the other dataset. Left/right results are SNR/LSD.}
\label{tab:out_of_distribution}
\begin{tabular}{l|c|c|}
\cline{2-3}
\multicolumn{1}{l|}{}                    & \multicolumn{1}{c|}{VCTK Multi (Test)} & \multicolumn{1}{c|}{Piano (Test)} \\ \hline
\multicolumn{1}{|l|}{VCTK Multi (Train)} & 20.0/1.7                               & 24.4/2.5                          \\ \hline
\multicolumn{1}{|l|}{Piano (Train)}      & 10.2/2.6                               & 25.7/1.5                          \\ \hline
\end{tabular}
\end{table}

\section{Discussion}
In this section, we present further analyses and discuss some important aspects of the audio super-resolution task.

\paragraph*{Computational performance.} In addition to capturing long-range dependencies, AFiLM trains faster. This is mainly due to the Transformer's ability to be more parallelizable \cite{vaswani2017attention}. We compare the run-time efficiency of TFiLM and AFiLM. As presented in Table \ref{tab:computational_per}, even though the number of parameters of AFiLM is higher, it trains on average over $34\%$ faster than the TFiLM model. Unlike the RNN-based model, AFiLM effiently exploits the GPU resulting in much faster training.

\paragraph*{Generalization of the super-resolution model.} We investigate the model's ability to generalize to other domains. In order to do this, we switch from speech to music and the other way around. The results are presented in Table \ref{tab:out_of_distribution}. As seen in previous models \cite{kuleshov2017audio,birnbaum2019temporal}, the super-resolved samples contain the high frequency details but still sound noisy. The models specialize to the specific type of audio they are trained on. 

\section{Conclusion}
\label{sec:conclusion}
In this work, we introduce the use of self-attention for the audio super-resolution task. We present the Attention-based Feature-Wise Linear Modulation (AFiLM) layer which relies on attention instead of recurrent neural networks to alter the activations of the convolutional model. The resulting model efficiently captures long-range temporal interactions. It outperforms all previous models and can be trained faster.

In future work, we want to develop super-resolution models that generalize well to different types of inputs. We also want to investigate perceptual-based models.

\section{Acknowledgements}
We would like to thank Bruce Basset for his helpful comments and advice.

\bibliographystyle{IEEEbib}
\bibliography{refs}

\end{document}